\documentclass[10pt,twocolumn,twoside]{IEEEtran}

\usepackage{calc}

\usepackage{multirow}
\usepackage{cite}
\usepackage{graphicx,subfigure}
\usepackage{psfrag}
\usepackage{amsmath,amssymb}
\usepackage{color}

\DeclareMathOperator*{\defeq}{\triangleq}

\newtheorem{theorem}{Theorem}

\newtheorem{proposition}{Proposition}
\newtheorem{example}{Example}

\newcommand{\bit}{\begin{itemize}}
\newcommand{\eit}{\end{itemize}}

\newcommand{\bc}{\begin{center}}
\newcommand{\ec}{\end{center}}

\newcommand{\ba}{\begin{array}}
\newcommand{\ea}{\end{array}}

\newcommand{\beq}{\begin{equation}}
\newcommand{\eeq}{\end{equation}}

\newcommand{\beqn}{\begin{equation*}}
\newcommand{\eeqn}{\end{equation*}}

\newcommand{\bean}{\begin{eqnarray*}}
\newcommand{\eean}{\end{eqnarray*}}
\newcommand{\bea}{\begin{eqnarray}}
\newcommand{\eea}{\end{eqnarray}}

\def\E{\mathbb{E}}

\def\ev{\boldsymbol{e}}

\def\gv{\boldsymbol{g}}
\def\hv{\boldsymbol{h}}

\def\wv{\boldsymbol{w}}
\def\xv{\boldsymbol{x}}

\newcommand{\Dc}{{\mathcal D}}

\newcommand{\Xc}{{\mathcal X}}

\newcommand{\T}{{\scriptscriptstyle\mathsf{T}}}
\renewcommand{\H}{{\scriptscriptstyle\mathsf{H}}}

\newtheorem{remark}{Remark}

\usepackage{mathtools}
\usepackage{xifthen}

\newcommand{\akt}[1][]{\ifthenelse{\isempty{#1}}{\alpha_{k,t}}{\alpha_{#1,t}}}
\newcommand{\aavg}[1][]{\ifthenelse{\isempty{#1}}{\bar{\alpha}}{\bar{\alpha}_{#1}}}
\newcommand{\FBfracP}[1][]{\ifthenelse{\isempty{#1}}{\delta_{\textrm{P}}}{\delta_{\textrm{P},#1}}}
\newcommand{\FBfracPv}[1][]{\ifthenelse{\isempty{#1}}{\boldsymbol{\delta}_{\textrm{P}}}{\boldsymbol{\delta}_{\textrm{P},#1}}}
\newcommand{\FBfracD}[1][]{\ifthenelse{\isempty{#1}}{\delta_{\textrm{D}}}{\delta_{\textrm{D},#1}}}

\begin{document}
\sloppy
\title{Optimal DoF Region of the Two-User MISO-BC with General Alternating CSIT}
\author{Jinyuan Chen and Petros Elia
\thanks{The research leading to these results has received funding from the European Research Council under the European Community's Seventh Framework Programme (FP7/2007-2013) / ERC grant agreement no. 257616 (CONECT), from the FP7 CELTIC SPECTRA project, and from Agence Nationale de la Recherche project ANR-IMAGENET.
}
\thanks{J. Chen and P. Elia are with the Mobile Communications Department, EURECOM, Sophia Antipolis, France (email: \{chenji, elia\}@eurecom.fr). }
}

\maketitle
\thispagestyle{empty}

\begin{abstract}
In the setting of the time-selective two-user multiple-input single-output (MISO) broadcast channel (BC), recent work by Tandon et al. considered the case where - in the presence of error-free delayed channel state information at the transmitter (delayed CSIT) - the current CSIT for the channel of user 1 and of user 2, alternate between the two extreme states of perfect current CSIT and of no current CSIT.

Motivated by the problem of having limited-capacity feedback links which may not allow for perfect CSIT, as well as by the need to utilize any available partial CSIT, we here deviate from this `all-or-nothing' approach and proceed - again in the presence of error-free delayed CSIT - to consider the general setting where current CSIT now alternates between any two qualities.  Specifically for $I_1$ and $I_2$ denoting the high-SNR asymptotic rates-of-decay of the mean-square error of the CSIT estimates for the channel of user~1 and of user~2 respectively, we consider the case where $I_1,I_2 \in\{\gamma,\alpha\}$ for any two positive current-CSIT quality exponents $\gamma,\alpha$; as a result, the overall current CSIT - for both users' channels - alternates between any four states $I_1I_2\in\{\gamma\gamma,\gamma\alpha,\alpha\gamma,\alpha\alpha\}$.  In a fast-fading setting where we consider communication over any number of coherence periods, and where each CSIT state $I_1I_2$ is
  present for a fraction $\lambda_{I_1I_2}$ of this total duration (naturally forcing $\lambda_{\alpha\gamma}+\lambda_{\gamma\alpha}+\lambda_{\alpha\alpha}+\lambda_{\gamma\gamma}=1$), we focus on the symmetric case of $\lambda_{\alpha\gamma}=\lambda_{\gamma\alpha}$, and derive the optimal degrees-of-freedom (DoF) region to be the polygon with corner points $\{(0,0), (0,1), (\bar{\lambda},1), (\frac{2+\bar{\lambda}}{3},\frac{2+\bar{\lambda}}{3}),(1,\bar{\lambda}),(1, 0)\}$, for some $ \bar{\lambda} \defeq (\lambda_{\gamma\alpha} + \lambda_{\gamma\gamma})\gamma +(\lambda_{\alpha\gamma}+\lambda_{\alpha\alpha})\alpha,$ representing a measure of the average CSIT quality.
The result, which is supported by novel communication protocols, naturally incorporates the aforementioned `Perfect current' vs. `No current' setting by limiting $I_1,I_2\in\{0,1\}$, as well as the Yang et al. and Gou and Jafar setting by forcing $\alpha = \gamma$.

Finally, motivated by recent interest in frequency correlated channels with unmatched CSIT, we also analyze the setting where there is no delayed CSIT.
\end{abstract}

\section{Introduction}

This work considers the two-user ($K=2$) $M$-transmit antenna $(M\geq K)$ multiple-input single-output (MISO) broadcast channel (BC) which accepts the input-output channel model
\begin{subequations}
\begin{align}
y^{(1)}_t &= \hv^{\T}_t \xv_t + z^{(1)}_t      \label{eq:modely1}\\
y^{(2)}_t &= \gv^{\T}_t \xv_t + z^{(2)}_t      \label{eq:modely2}
\end{align}
\end{subequations}
where $y^{(k)}_t$ denotes the received signal of user-$k$ during time-slot $t$, $\hv_t, \gv_t $ denote the $M\times 1$ channel vectors, $z^{(k)}_t$ denotes the unit power AWGN noise, and where $\xv_{t}$ represents the transmitted signal vector adhering to a power constraint $\E[ ||\xv_{t}||^2 ] \le P$, with $P$ also taking the role of the signal-to-noise ratio (SNR).
Corresponding to the fast fading case, the coefficients $\hv_{t}, \gv_t $ are modeled as independent and identically distributed (i.i.d.) complex Gaussian random variables with zero mean and unit variance.

In this setting, the performance is heavily affected by the timeliness and quality of the channel state information at the transmitter (CSIT); as is well known, having full CSIT allows for the optimal 1 degrees-of-freedom (DoF) per user (cf.~\cite{CS:03})\footnote{We remind the reader that for an achievable rate tuple $(R_1,R_2)$, where $R_k$ is for user~$k$, the corresponding DoF tuple $(d_1,d_2)$ is given by $d_i = \lim_{P \to \infty} \frac{R_i}{\log P},\ i=1,2$.  The corresponding DoF region $\Dc$ is then the set of all achievable DoF tuples $(d_1,d_2)$.}, while having no CSIT allows only for $1/2$ DoF per user (cf.~\cite{JG:05,HJSV:12}).
This significant gap has spurred research efforts to analyze and optimize communications in the presence of delayed and imperfect feedback.  One important contribution came with the work of Maddah-Ali and Tse in \cite{MAT:11c} which revealed the benefits of employing delayed CSIT even if this CSIT is completely obsolete.  In a setting that differentiated between current and \emph{delayed CSIT} - delayed CSIT being that which is available after the channel elapses, i.e., after the end of the coherence period corresponding to the channel described by this delayed feedback, while current CSIT corresponding to feedback received during the channel's coherence period - the work in \cite{MAT:11c} showed that perfect delayed CSIT, even without any current CSIT, allows for an improved $2/3$ DoF per user.
Several interesting generalizations followed, including the work in~\cite{YKGY:12d,GJ:12o,CE:12d} which explored the setting of combining perfect delayed CSIT with immediately available (current) imperfect (partial) CSIT, the work in \cite{CE:12c} which additionally considered the effects of the quality of delayed CSIT,  and the work in \cite{CE:12m} which considered delayed and progressively evolving current CSIT.  Other interesting works in the context of utilizing delayed and current CSIT, can be found for example in \cite{VV:11t,GMK:11o,AGK:11o,GMK:11i,XAJ:11b,LSW:12,TMT+:12,LH:12,TJS:12}.

An interesting generalization of the delayed CSIT setting in \cite{MAT:11c} - and a starting point of our work here - came with the work by Tandon et al. in \cite{TJSP:12} which, for the setting of the time-selective two-user MISO BC, considered the case where - in the presence of error-free delayed CSIT\footnote{We clarify that, while \cite{TJSP:12} allowed for the possibility that delayed CSIT may or may not be present, we are here assuming that delayed CSIT is indeed available.} - the current CSIT for the channel of user 1 and of user 2, alternates between the two extreme states of perfect current CSIT and of no current CSIT.

\subsection{CSIT quantification and the any-two-state alternating feedback model}

Under the assumption of constantly available error-free delayed CSIT, we draw from the alternating CSIT setting, and consider here the more general setting where current CSIT - for each user's channel - now alternates between \emph{any} two qualities.
Specifically for $\hat{\hv}_{t}, \hat{\gv}_{t}$ denoting the current CSIT estimates for channels $\hv_{t}, \gv_{t}$ respectively, for
\[\tilde{\hv}_{t}= \hv_{t} - \hat{\hv}_{t} , \    \tilde{\gv}_{t}= \gv_{t} - \hat{\gv}_{t}  \]
denoting estimation errors, each mutually independent of the estimates, and each having i.i.d entries, and for $I_1$ and $I_2$
\[I_1\defeq -\frac{\log \E\bigl[\|\tilde{\hv}_{t}\|^2\bigr]}{ \log P} , \quad  I_2\defeq -\frac{\log \E\bigl[\|\tilde{\gv}_{t}\|^2\bigr]}{ \log P}  \]
denoting the high-SNR asymptotic rates-of-decay of the mean-square error of the CSIT estimates for the channel of user~1 and of user~2 respectively, we consider the case where \[I_1,I_2 \in\{\gamma,\alpha\}\] for any two positive current-CSIT quality exponents $\gamma,\alpha$.
We note that in the DoF setting of interest, and without loss of generality, these exponents can be bounded as \[0\leq \alpha \leq \gamma \leq 1\] where $\alpha  = 0$ (or $\gamma  = 0$) implies no (or very little) current CSIT knowledge, and where $\alpha  = 1$ (or $\gamma  = 1$) implies essentially perfect CSIT (cf. \cite{Caire+:10m}).
Furthermore noting that the overall current CSIT - for both users' channels - alternates between any four states \[I_1I_2\in\{\gamma\gamma,\gamma\alpha,\alpha\gamma,\alpha\alpha\}\] we consider the case where each joint CSIT state $I_1I_2$ is present for a fraction $\lambda_{I_1I_2}$ of the total communication duration. Finally, as in \cite{TJSP:12}, we are interested in the symmetric case where \[\lambda_{\alpha\gamma}=\lambda_{\gamma\alpha}.\]

Our setting and generalization is naturally motivated by the fact that finite-capacity feedback links may never allow for perfect CSIT, but instead may allow for CSIT estimates that, albeit imperfect, can still be useful.  Interest in analyzing and encoding in the presence of partial-CSIT, again comes from the use of limited-capacity feedback links, as well as from possible channel correlations in time and/or frequency; see for example the work in~\cite{YKGY:12d,GJ:12o}, as well as the work in \cite{HC:13} which considers a frequency-correlated setting where CSIT estimates can be partially extrapolated between adjacent frequency sub-bands.

\subsection{Notation and conventions}

We will henceforth consider a fast fading channel representation where our time index $t$ is normalized so that channel realizations change - from one channel use to another - in an i.i.d manner\footnote{Note that the i.i.d. assumption only affects the DoF outer bounds, and that the schemes indeed achieve the described DoF performance even in the presence of channel correlation.}. Furthermore, as is common, we will consider perfect and global knowledge of channel state information at the receivers, as well as will allow each receiver to perfectly know all CSI and all CSIT estimates.

In terms of notation, $(\bullet)^\T$, $(\bullet)^{\H}$ will denote the transpose and conjugate transpose of a matrix respectively, while $||\bullet||$ will denote the Euclidean norm, and $|\bullet|$ will denote either the magnitude of a scalar or the cardinality of a set.
$\ev^{\bot}$ will denote a unit-norm vector orthogonal to $\ev$.
$o(\bullet)$ comes from the standard Landau notation, where $f(x) = o(g(x))$ implies $\lim_{x\to \infty} f(x)/g(x)=0$.
We will also use $\doteq$ to denote \emph{exponential equality}, i.e., we will write $f(P)\doteq P^{B}$ to denote $\displaystyle\lim_{P\to\infty}\frac{\log f(P)}{\log P}=B$.
Logarithms are of base~$2$.

\section{DoF Region of Two-User MISO-BC with Any-Two-State Alternating Current CSIT\label{sec:bc-dof}}

We proceed to describe in Theorem~\ref{thm:alterFre} the optimal DoF region of the MISO BC with any-two-state alternating current CSIT and perfect delayed CSIT, while after that we give DoF bounds for the case where there is no delayed CSIT. Finally in Section~\ref{sec:schemes} we describe the new precoding protocols that achieve the corresponding DoF corner points.
For notational convenience, we let \[ \bar{\lambda} \defeq (\lambda_{\gamma\alpha} + \lambda_{\gamma\gamma})\gamma +(\lambda_{\alpha\gamma}+\lambda_{\alpha\alpha})\alpha\] which can be readily interpreted as an average measure of current CSIT quality.

\subsection{Any-two-state alternating current CSIT with perfect delayed CSIT}
	
\vspace{2pt}
\begin{theorem}\label{thm:alterFre}
For the two-user MISO BC with alternating current-CSIT quality-exponents $\alpha, \gamma$, and given perfect delayed CSIT, the optimal DoF region is
  \begin{align} \label{eq:alterFre}
	   d_1 \le 1, \quad  \  d_2 \le 1   \\
     2d_1  + d_2 \le 2 + \bar{\lambda} \\
     2d_2  + d_1 \le 2 + \bar{\lambda}
  \end{align}
and corresponds to the polygon with corner points \[\{(0,0), (0,1), (\bar{\lambda},1), (\frac{2+\bar{\lambda}}{3},\frac{2+\bar{\lambda}}{3}),(1,\bar{\lambda}),(1, 0)\}.\]
\end{theorem}
\vspace{2pt}
\begin{proof}
The converse part of the proof is derived directly from~\cite{CYE:13isit}, while achievability is shown in Section~\ref{sec:schemes}.
\end{proof}

\begin{remark}
As noted, the derived region incorporates - under the assumption of constantly available delayed CSIT - the result in \cite{TJSP:12} for the special case where $\gamma=1, \alpha=0$, as well as the result in \cite{YKGY:12d,GJ:12o} for the special case where $\gamma= \alpha$.
\end{remark}

\begin{example}
For $\lambda_{\alpha\gamma} = \lambda_{\gamma\alpha} = 1/2$, the alternating pattern could have the form
\[\ba{c|ccccc}
t             & 1          & 2                &  3         & 4  &   \cdots    \\
I_1 & \gamma   & \alpha         & \gamma   & \alpha &\\
I_2  & \alpha   & \gamma         & \alpha   & \gamma &\ea
\]
while for $\lambda_{\gamma\alpha}=\lambda_{\alpha\gamma}= \lambda_{\gamma\gamma}= \lambda_{\alpha\alpha}= \frac{1}{4}$, the alternating pattern could have the form
\[\ba{c|ccccc}
t             & 1          & 2                &  3         & 4     &\cdots    \\
I_1 & \gamma   & \alpha         & \gamma   & \alpha & \\
I_2  & \alpha   & \gamma         & \gamma & \alpha. & \ea
\]
For any $\alpha,\gamma$ such that $\alpha+\gamma = 1$, both cases would allow for a symmetric DoF of $d_1 = d_2 = \frac{4+\alpha+\gamma}{6} = \frac{5}{6}$, and again both cases have as special instances, the $\alpha = 0,\gamma = 1$ case corresponding to~\cite{TJSP:12}, and the instance of $\alpha = \gamma = 1/2$ from \cite{YKGY:12d,GJ:12o}.
\end{example}

\subsection{Any-two-state alternating current CSIT, with no delayed CSIT}
We here consider the previous scenario, without though any delayed CSIT. The following proposition provides an inner bound, while the result of Theorem~\ref{thm:alterFre} naturally serves as an outer bound.
\vspace{2pt}
\begin{proposition}\label{prop:alterFreNodelay}
For the two-user MISO BC with alternating current-CSIT quality-exponents $\alpha, \gamma$, and no delayed CSIT, the DoF region
  \begin{align} \label{eq:alterFreNod}
	   d_1 \le 1, \quad   d_2 \le 1   \\
     d_1  + d_2 \le 1 + \bar{\lambda}
  \end{align}
is achievable and it corresponds to a polygon with corner points \[\{(0,0), (0,1), (\bar{\lambda},1),(1,\bar{\lambda}),(1, 0)\}.\]
\end{proposition}
\vspace{2pt}
\begin{proof}
The proof is direct from the schemes proposed in Section~\ref{sec:schemes}.
\end{proof}

We proceed with an example inspired by the recently proposed setting (cf.~\cite{HC:13}) of the frequency correlated channel with unmatched CSIT, where CSIT estimates can be partially extrapolated between adjacent frequency sub-bands.

\begin{example}
For the case where $\lambda_{\gamma\alpha}=\lambda_{\alpha\gamma}=\frac{1}{2}$,
corresponding to a setting where
\[\ba{c|cc}
             & \text{sub-band} \ 1          & \text{sub-band} \ 2   \\
\text{user 1:} \ I_1 & \gamma   & \alpha   \\
\text{user 2:} \ I_2  & \alpha   & \gamma
\ea
\]
then the achievable DoF region described in proposition \ref{prop:alterFreNodelay}, takes the form
of a polygon with corner points \[\{(0,0), (0,1), ((\alpha+\gamma)/2,1),(1,(\alpha+\gamma)/2),(1, 0)\}\]
which corresponds to a symmetric DoF point of
\[d_1 = d_2 = \frac{2+\alpha+\gamma}{4}\] for any $\alpha,\gamma\in[0,1]$.
\end{example}

\begin{figure}
	\centering
	\includegraphics[width = 9cm]{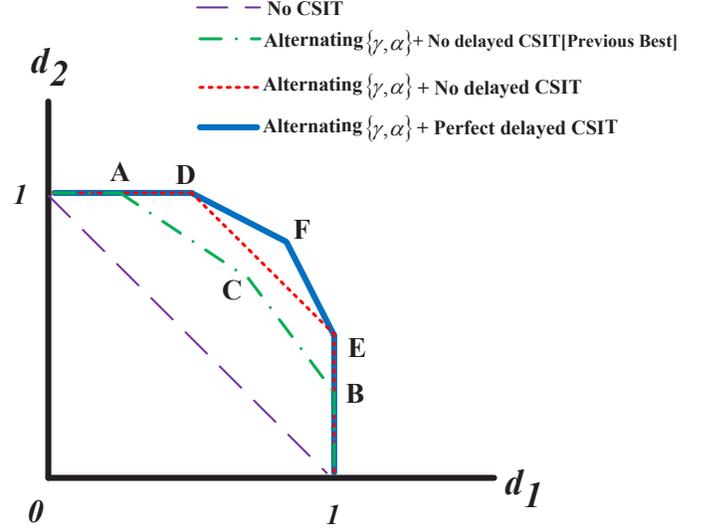}
	\caption{DoF regions of the two-user MISO BC with  two state $\{\gamma, \alpha\}$ alternating current CSIT, for case with $\gamma > \frac{2+\alpha}{3}$, where $A=(\alpha,1)$, $B=(1, \alpha)$, $C=(\frac{2+\alpha}{3}, \frac{2+\alpha}{3})$, $D=(\bar{\lambda}, 1)$, $E=(1, \bar{\lambda})$, $F=(\frac{2+\bar{\lambda}}{3} , \frac{2+\bar{\lambda}}{3})$. }
	\label{fig:nodcsit}
\end{figure}

\section{Communication schemes for the MISO BC with  any-two-state alternating current CSIT\label{sec:schemes}}
We proceed to describe the communication schemes that achieve the corresponding DoF corner points, by properly utilizing different combinations of superposition coding, successive interference cancellation, power allocation, and phase durations.

We will describe schemes that achieve specific DoF corner points for the specific cases of $\lambda_{\gamma\alpha} = \lambda_{\alpha\gamma} = 1/2$ and $\lambda_{\gamma\gamma}=1$ and $\lambda_{\alpha\alpha} = 1$, and we will then describe how to combine these schemes to achieve any DoF point for any desired $\lambda_{\gamma\gamma},\lambda_{\alpha\alpha},\lambda_{\gamma\alpha} = \lambda_{\alpha\gamma}$.

Specifically scheme $\Xc_{1}$ will require delayed CSIT and it will correspond to $\lambda_{\gamma\alpha} = \lambda_{\alpha\gamma} = 1/2$, scheme $\Xc_2$ will again consider $\lambda_{\gamma\alpha} = \lambda_{\alpha\gamma} = 1/2$ and no delayed CSIT, while schemes $\Xc_3,\Xc_4$ will be described for $\lambda_{\alpha\alpha} = 1$ and $\lambda_{\gamma\gamma} = 1$, with $\Xc_3$ requiring no delayed CSIT, while $\Xc_4$  - which is directly drawn from~\cite{YKGY:12d,GJ:12o} - requires delayed CSIT, and will achieve DoF point $(\frac{2+\gamma}{3}, \frac{2+\gamma}{3})$ for $\lambda_{\gamma\gamma}=1$, and DoF point $(\frac{2+\alpha}{3}, \frac{2+\alpha}{3})$ with $\lambda_{\alpha\alpha}=1$.

In terms of notation that is specific to the schemes, for any symbol $x_t$, we will use $P^{(x)}_{t} \defeq \E |x_{t}|^2$ to denote the power, and we will use $r^{(x)}_{t}$ to denote the prelog factor of the number of bits $r^{(x)}_{t}\log P - o(\log P)$ carried by $x_{t}$.
We will also use $\iota^{(1)}_t,\iota^{(2)}_t$ to respectively define the interference experienced by the first and second user at time $t$, we will use $\check{\iota}^{(1)}_t,\check{\iota}^{(2)}_t$ to denote the delayed estimates of this interference (at the transmitter, using delayed CSIT), and we will use $\bar{\check{\iota}}^{(1)}_t,\bar{\check{\iota}}^{(2)}_t$ to denote a quantized version of these estimates.
Furthermore in the setting where we quantize a set $x$ of complex numbers, we will use $\phi(x)$ to mean that the corresponding number of quantization bits is $\phi(x)\log P$.
Also, in describing schemes where communication is divided in phases, we will adopt a double time index $(s,t)$ representing time-slot $t$ of phase $s$.  Finally, for brevity of scheme description, we will often ignore the noise terms where they do not affect the DoF performance.

\subsection{Scheme $\Xc_{1}$: achieving DoF points $(\frac{4+\gamma+\alpha}{6}, \frac{4+\gamma+\alpha}{6})$ with $\lambda_{\gamma\alpha}=\lambda_{\alpha\gamma}=\frac{1}{2}$, and with delayed CSIT}

Scheme $\Xc_{1}$ will be a concatenation of two sub-schemes $\Xc^{'}_{1}$ and $\Xc^{''}_{1}$, where $\Xc^{''}_{1}$ is simply a reordered version of $\Xc^{'}_{1}$.

$\Xc^{'}_{1}$ has four phases with durations $T_1, T_2, T_3, T_4$ that are chosen as integers\footnote{$\alpha,\gamma$ are assumed to be rational numbers.} such that
\begin{align} \label{eq:T1T2}
T_2 =T_3=T_4 = T_1 \frac{2-\gamma-\alpha}{3(1-\gamma)}. \end{align}

While our approach in designing this scheme holds for any $I_1I_2$ CSIT pattern that satisfies $\lambda_{\gamma\alpha}=\lambda_{\alpha\gamma}=\frac{1}{2}$, without loss of generality we will assume the alternating pattern where the joint state $I_1I_2$ continuously alternates between $I_1I_2 = \alpha\gamma$ and $I_1I_2 = \gamma\alpha$, as suggested in Table~\ref{tab:x1half}.  As commented in~\cite{TJSP:12}, for a sufficiently large communication duration, this assumption introduces no restrictions.
\begin{table}
\caption{Current CSIT alternating pattern for $\Xc^{'}_{1}$.}
\begin{center}
\begin{tabular}{|c|c|c|c|c|}
  \hline
Phase                               & 1        & 2          & 3          & 4          \\
  \hline
Duration                               & $T_1$        & $T_2$           & $T_3$           & $T_4$           \\
  \hline
$I_1$       & $\alpha$ & $\gamma$   & $\alpha$   & $\gamma$       \\
   \hline
$I_2$        & $\gamma$ & $\alpha$   & $\gamma$   & $\alpha$    \\
   \hline
\end{tabular}
\end{center}
\label{tab:x1half}
\end{table}
We proceed with the description of the phases.

\paragraph{Phase~1, $(s=1, t=1,\cdots, T_1, I_1=\alpha, I_2=\gamma)$}
\begin{enumerate}
\item During phase~1, the transmitter sends
\beq \label{eq:TxX11t1}
\xv_{1,t} = \hat{\gv}^{\bot}_{1,t} a_{1,t} + \hat{\hv}_{1,t} a^{'}_{1,t}+ \hat{\hv}^{\bot}_{1,t} b_{1,t} + \hat{\gv}_{1,t} b^{'}_{1,t} \eeq
with
\begin{equation}\label{eq:rpX11t1}
\begin{array}{cccc}
 P^{(a)}_{1,t} \!\doteq\! P, \!\!& P^{(a')}_{1,t} \!\doteq\!  P^{1-\gamma}, \!\!& P^{(b)}_{1,t} \!\doteq\! P, \!\!& P^{(b')}_{1,t} \!\doteq\! P^{1- \alpha}\\
 r^{(a)}_{1,t}  \!=\! 1,     \!\!& r^{(a')}_{1,t}  \!=\!  1-\gamma,       \!\!& r^{(b)}_{1,t} \!=\!1,         \!\!& r^{(b')}_{1,t} \!=\! 1-\alpha. \end{array} \end{equation}
(private symbols $a_{1,t}, a^{'}_{1,t}$ for user~1, and $b_{1,t}, b^{'}_{1,t}$ for user~2)
\item At end of phase~1, the transmitter reconstructs delayed estimates $\{\check{\iota}^{(1)}_{1,t}, \check{\iota}^{(2)}_{1,t}\}_{t=1}^{T_1}$ of the interference $\iota^{(1)}_{1,t}\defeq \tilde{\hv}^\T_{1,t} \hat{\hv}^{\bot}_{1,t} b_{1,t}+\hv^\T_{1,t} \hat{\gv}_{1,t} b^{'}_{1,t}$, $\iota^{(2)}_{1,t}\defeq \tilde{\gv}^\T_{1,t}\hat{\gv}^{\bot}_{1,t} a_{1,t}+\gv^\T_{1,t} \hat{\hv}_{1,t} a^{'}_{1,t}$ at the first and second user respectively
\item Quantizes $\{\check{\iota}^{(1)}_{1,t}, \check{\iota}^{(2)}_{1,t}\}_{t=1}^{T_1}$ into $\{\bar{\check{\iota}}^{(1)}_{1,t},\bar{\check{\iota}}^{(2)}_{1,t}\}_{t=1}^{T_1}$, with quantization rate $\phi(\bar{\check{\iota}}^{(1)}_{1,t}) = 1-\alpha,   \quad   \phi(\bar{\check{\iota}}^{(2)}_{1,t}) = 1-\gamma $, to allow for bounded quantization noise power (cf. \cite{CT:06})
\item Evenly maps $T_1 (2 - \gamma - \alpha)\log P$ bits of $\{\bar{\check{\iota}}^{(2)}_{1,t},\bar{\check{\iota}}^{(1)}_{1,t}\}$ into set $\{c_{2,t},c_{3,t},c_{4,t}\}_{t=1}^{ T_2} $ \\ ($\{c_{2,t},c_{3,t},c_{4,t}\}_{t=1}^{ T_2}  $ will be sequentially transmitted in the next phases, in order to cancel interference, and serve as extra observations for decoding the private symbols in phase~1)
\end{enumerate}

\paragraph{Phase~2, $(s=2, t=1,\cdots, T_2, I_1=\gamma, I_2=\alpha)$}
\begin{enumerate}
\item Transmits
\beq \label{eq:TxX11t2}
\xv_{2,t} \!=\! \wv_{2,t} c_{2,t} +  \hat{\gv}^{\bot}_{2,t} ( a_{2,t} + a^{'}_{2,t}   ) + \hat{\hv}^{\bot}_{2,t}  a^{''}_{2,t} +  \hat{\hv}^{\bot}_{2,t} b_{2,t}  \eeq
with
\begin{equation}\label{eq:rpX11t2}
\begin{array}{ccccc}
\!\!\! P^{(c)}_{2,t} \!\!\doteq\!\! P, \!\!\!&\!\!\! P^{(a)}_{2,t} \!\!\doteq\!\! P^{\gamma}, \!&\!\! \!\!P^{(a')}_{2,t} \!\doteq\! P^{\alpha}, \!\!\!& \!\! P^{(a'')}_{2,t}\!\!\doteq\!\! P^{\gamma-\alpha}\!, \!\!&\!\!\! P^{(b)}_{2,t} \!\!\doteq\!\! P^{\gamma}\\
\!\!\! r^{(c)}_{2,t}  \!\!=\!\! 1\!\!-\!\!\gamma, \!\!\!\!&\! r^{(a)}_{2,t}  \!\!=\!\! \gamma\!-\!\!\alpha, \!\!\!\!& r^{(a')}_{2,t}  \!=\!  \alpha,     \!\!\!\!& r^{(a'')}_{2,t} \!\!=\!\!\gamma\!-\!\! \alpha,         \!\!\!\!&\!\!\! r^{(b)}_{2,t} \!\!=\! \gamma \end{array} \end{equation}
($a_{2,t}, a^{'}_{2,t}, a^{''}_{2,t}$ for user~1, $b_{2,t}$ for user~2, $c_{2,t}$ is common)
\item At end of phase~2, the transmitter reconstructs $\iota^{(2)}_{2,t}\defeq \tilde{\gv}^\T_{2,t}\hat{\gv}^{\bot}_{2,t} a_{2,t}+ \gv^\T_{2,t} \hat{\hv}^{\bot}_{2,t} a^{''}_{2,t}$ into $\check{\iota}^{(2)}_{2,t}, \ t=1,\cdots,T_2$
\item Quantizes $\check{\iota}^{(2)}_{2,t}$ into $\bar{\check{\iota}}^{(2)}_{2,t}$ with quantization rate $\phi(\bar{\check{\iota}}^{(2)}_{2,t}) =  \gamma-\alpha$
\item Maps $\bar{\check{\iota}}^{(2)}_{2,t}$ ($t=1,\cdots,T_2$), into $c^{'}_{3,t}$ for transmission over the next phase.
\end{enumerate}

\paragraph{Phase~3, $(s=3, t=1,\cdots, T_3, I_1=\alpha, I_2=\gamma)$}
\begin{enumerate}
\item Transmits
\beq \label{eq:TxX11t3}
\xv_{3,t} = \wv_{3,t} c_{3,t}  +  \hat{\gv}^{\bot}_{3,t} a_{3,t}  +  \hat{\gv}_{3,t} c^{'}_{3,t} +  \hat{\hv}^{\bot}_{3,t} b_{3,t}  \eeq
with
\begin{equation}\label{eq:rpX11t3}
\begin{array}{cccc}
 P^{(c)}_{3,t} \!\doteq \!P,   \!\! & P^{(a)}_{3,t} \!\doteq\! P^{\gamma}, \!\!& P^{(c^{'})}_{3,t} \!\doteq\! P^{\gamma},   \!\!& P^{(b)}_{3,t} \!\doteq\! P^{\alpha}\\
 r^{(c)}_{3,t}  \!=\! 1-\gamma, \!\!& r^{(a)}_{3,t}  \!=\! \gamma,         \!\!& r^{(c^{'})}_{3,t}  \!=\!  \gamma- \alpha,  \!\!& r^{(b)}_{3,t} \!=\! \alpha \end{array} \end{equation}
\end{enumerate}
\paragraph{Phase~4, $(s=4, t=1,\cdots, T_4, I_1=\gamma, I_2=\alpha)$}
\begin{enumerate}
\item Transmits
\beq \label{eq:TxX11t4}
\xv_{4,t} = \wv_{4,t} c_{4,t}  +  \hat{\gv}^{\bot}_{4,t} a_{4,t}  +  \hat{\gv}_{4,t} c^{'}_{3,t} +  \hat{\hv}^{\bot}_{4,t} b_{4,t}  \eeq
with
\begin{equation}\label{eq:rpX11t4}
\begin{array}{ccc}
 P^{(c)}_{4,t} \doteq P,    & P^{(a)}_{4,t} \doteq P^{\alpha},   & P^{(b)}_{4,t} \doteq P^{\gamma}\\
 r^{(c)}_{4,t}  = 1-\gamma, & r^{(a)}_{4,t}  = \alpha,          & r^{(b)}_{4,t} = \gamma \end{array} \end{equation}
($c^{'}_{3,t}$ is the same symbol sent in the previous phase)
\end{enumerate}
\vspace{4pt}
Moving on to the decoding part, we proceed to describe each step, taking into consideration the rates and powers in \eqref{eq:rpX11t1},\eqref{eq:rpX11t2},\eqref{eq:rpX11t3},\eqref{eq:rpX11t4}, as well as the nature of the (noiseless) signals described below for each of the four phases, where we also note the order of each summand's average power.

\begin{align}
  y^{(1)}_{1,t} &= \underbrace{\hv^\T_{1,t} \hat{\gv}^{\bot}_{1,t} a_{1,t}}_{P}+ \underbrace{\hv^\T_{1,t} \hat{\hv}_{1,t} a^{'}_{1,t}}_{P^{1-\gamma}} + \overbrace{\underbrace{\tilde{\hv}^\T_{1,t} \hat{\hv}^{\bot}_{1,t} b_{1,t}}_{P^{1-\alpha}} +\underbrace{\hv^\T_{1,t} \hat{\gv}_{1,t} b^{'}_{1,t}}_{P^{1-\alpha}}}^{\iota^{(1)}_{1,t}} \label{eq:x11t1y1}\\
  y^{(2)}_{1,t} &= \overbrace{\underbrace{\tilde{\gv}^\T_{1,t}\hat{\gv}^{\bot}_{1,t} a_{1,t}}_{P^{1-\gamma}} + \underbrace{\gv^\T_{1,t} \hat{\hv}_{1,t} a^{'}_{1,t}}_{P^{1-\gamma}}}^{\iota^{(2)}_{1,t}} + \underbrace{\gv^\T_{1,t} \hat{\hv}^{\bot}_{1,t} b_{1,t}}_{P}+ \underbrace{\gv^\T_{1,t} \hat{\gv}_{1,t} b^{'}_{1,t}}_{P^{1-\alpha}} \label{eq:x11t1y2}
\end{align}


\begin{align}
  y^{(1)}_{2,t} \!\!&=\! \underbrace{\hv^\T_{2,t} \wv_{2,t} c_{2,t}}_{P} \!+\! \underbrace{\hv^\T_{2,t} \hat{\gv}^{\bot}_{2,t} a_{2,t}}_{P^{\gamma}}\!+\! \underbrace{\hv^\T_{2,t} \hat{\gv}^{\bot}_{2,t}  a^{'}_{2,t} }_{P^{\alpha}} \!+\! \underbrace{\tilde{\hv}^\T_{2,t} \hat{\hv}^{\bot}_{2,t} a^{''}_{2,t}}_{P^{-\alpha}} \nonumber\\ & \quad \!+\! \underbrace{\tilde{\hv}^\T_{2,t} \hat{\hv}^{\bot}_{2,t} b_{2,t}}_{P^{0}}  \label{eq:x11t2y1}\\
  y^{(2)}_{2,t} \!\!&=\! \underbrace{\gv^\T_{2,t} \wv_{2,t} c_{2,t}}_{P} \!+\! \overbrace{\underbrace{\tilde{\gv}^\T_{2,t}\hat{\gv}^{\bot}_{2,t} a_{2,t}}_{P^{ \gamma-\alpha}} \!+\! \underbrace{\gv^\T_{2,t} \hat{\hv}^{\bot}_{2,t} a^{''}_{2,t}}_{P^{ \gamma-\alpha}}}^{\iota^{(2)}_{2,t}} \!+\! \underbrace{\tilde{\gv}^\T_{2,t}\hat{\gv}^{\bot}_{2,t} a^{'}_{2,t}}_{P^{0}} \nonumber\\ & \quad \!+\! \underbrace{\gv^\T_{2,t} \hat{\hv}^{\bot}_{2,t} b_{2,t}}_{P^{\gamma}} \label{eq:x11t2y2}
\end{align}

\begin{align}
  y^{(1)}_{3,t} &= \underbrace{\hv^\T_{3,t} \wv_{3,t} c_{3,t}}_{P} + \underbrace{\hv^\T_{3,t} \hat{\gv}^{\bot}_{3,t} a_{3,t}}_{P^{\gamma}} + \underbrace{\hv^\T_{3,t} \hat{\gv}_{3,t} c^{'}_{3,t} }_{P^{\gamma}} + \underbrace{\tilde{\hv}^\T_{3,t} \hat{\hv}^{\bot}_{3,t} b_{3,t}}_{P^{0}} \label{eq:x11t3y1}\\
  y^{(2)}_{3,t} &= \underbrace{\gv^\T_{3,t} \wv_{3,t} c_{3,t}}_{P} + \underbrace{\tilde{\gv}^\T_{3,t}\hat{\gv}^{\bot}_{3,t} a_{3,t}}_{P^{0}} + \underbrace{\gv^\T_{3,t}\hat{\gv}_{3,t} c^{'}_{3,t}}_{P^{\gamma}}+ \underbrace{\gv^\T_{3,t} \hat{\hv}^{\bot}_{3,t} b_{3,t}}_{P^{\alpha}} \label{eq:x11t3y2}
\end{align}

\begin{align}
  y^{(1)}_{4,t} &= \underbrace{\hv^\T_{4,t} \wv_{4,t} c_{4,t}}_{P} + \underbrace{\hv^\T_{4,t} \hat{\gv}^{\bot}_{4,t} a_{4,t}}_{P^{\alpha}}  + \underbrace{\hv^\T_{4,t} \hat{\gv}_{4,t} c^{'}_{3,t} }_{P^{\gamma}} + \underbrace{\tilde{\hv}^\T_{4,t} \hat{\hv}^{\bot}_{4,t} b_{4,t}}_{P^{0}}  \label{eq:x11t4y1}\\
  y^{(2)}_{4,t} &= \underbrace{\gv^\T_{4,t} \wv_{4,t} c_{4,t}}_{P} + \underbrace{\tilde{\gv}^\T_{4,t}\hat{\gv}^{\bot}_{4,t} a_{4,t}}_{P^{0}} + \underbrace{\gv^\T_{4,t}\hat{\gv}_{4,t} c^{'}_{3,t}}_{P^{\gamma}}+ \underbrace{\gv^\T_{4,t} \hat{\hv}^{\bot}_{4,t} b_{4,t}}_{P^{\gamma}}. \label{eq:x11t4y2}
\end{align}
We continue with the description of the decoding.\\[5pt]
Decoding for user~1 for phase $2,3,4$:
\begin{enumerate}
\item Immediate decoding of $c_{2,t}, c_{3,t}, c_{4,t}$, treating other signals as noise
\item Goes back to phase~2, removes $c_{2,t}$ from $y^{(1)}_{2,t}$ in \eqref{eq:x11t2y1}, and decodes $a_{2,t},a^{'}_{2,t}$ using successive decoding (SD)
\item Goes to phase~4, removes $c_{4,t}$ from $y^{(1)}_{4,t}$ in~\eqref{eq:x11t4y1}, and decodes $c^{'}_{3,t}$ and $a_{4,t}$ using SD
\item Goes to phase~3, removes $c_{3,t}$ and $c^{'}_{3,t}$ from $y^{(1)}_{3,t}$ in~\eqref{eq:x11t3y1}, and directly decodes $a_{3,t}$
\item Goes back to phase~2, and directly decodes $a^{''}_{2,t}$ using the acquired knowledge of $a_{2,t}$ and of $c^{'}_{3,t}$ (i.e., of $\bar{\check{\iota}}^{(2)}_{2,t}$) (see~\eqref{eq:x11t2y2})
\end{enumerate}
Decoding for user~2 for phase $2,3,4$:
\begin{enumerate}
\item Immediate decoding of $c_{2,t}, c_{3,t}, c_{4,t}$
\item Goes back to phase $3$, removes $c_{3,t}$ from $y^{(2)}_{3,t}$ in \eqref{eq:x11t3y2}, and decodes $c^{'}_{3,t}$ and $b_{3,t}$ using SD
\item Goes to phase $4$, removes $c^{'}_{3,t}$ and $c_{4,t}$ from $y^{(2)}_{3,t}$ in \eqref{eq:x11t4y2}, and directly decodes $b_{4,t}$
\item Goes to phase $2$, removes $\bar{\check{\iota}}^{(2)}_{2,t}$ and $c_{2,t}$ from $y^{(2)}_{2,t}$ in \eqref{eq:x11t2y2}, and directly decodes $b_{2,t}$.
\end{enumerate}
Decoding for user~1 and user~2 for phase $1$:
\begin{enumerate}
\item Both users reconstruct $\{\bar{\check{\iota}}^{(1)}_{1,t},\bar{\check{\iota}}^{(2)}_{1,t}\}_{t=1}^{T_1}$ from  decoded $\{c_{2,t}, c_{3,t}, c_{4,t}\}_{t=1}^{T_2}$
\item User~1 removes $\bar{\check{\iota}}^{(1)}_{1,t}$ from $y^{(1)}_{1,t}$, uses $\bar{\check{\iota}}^{(2)}_{1,t}$ as a new observation, and decodes $a_{1,t}$ and $a^{'}_{1,t}$
\item User~2 removes $\bar{\check{\iota}}^{(2)}_{1,t}$ from $y^{(2)}_{1,t}$, uses $\bar{\check{\iota}}^{(1)}_{1,t}$ as a new observation, and decodes $b_{1,t}$ and $b^{'}_{1,t}$.
\end{enumerate}
\vspace{3pt}Adding up the bits throughout the four phases, gives a sum DoF of
\begin{align}
d_{\sum}&=d_1+d_2  = \frac{ T_1 ( 4- \gamma - \alpha) + T_2(5 \gamma+ \alpha)  }{T_1+3T_2}  \nonumber\\
&= \frac{ T_1 ( 4- \gamma - \alpha) + T_1 \frac{2-\gamma-\alpha}{3(1-\gamma)} (5 \gamma+ \alpha)  }{T_1+T_1 \frac{2-\gamma-\alpha}{(1-\gamma)}}  \nonumber\\
&= \frac{ 4+\gamma+ \alpha  }{3}. \label{eq:X11sumDoF}
\end{align}

In order to achieve the symmetric DoF  $d_1=d_2=\frac{ 4+\gamma+ \alpha  }{6}$ with $\lambda_{\gamma\alpha}=\lambda_{\alpha\gamma}=\frac{1}{2}$, we concatenate sub-scheme $\Xc^{'}_{1}$ with its reordered version $\Xc^{''}_{1}$ which corresponds to the CSIT alternating sequence suggested in Table~\ref{tab:x1half2}, and for which we interchange the $a$ and $b$ symbols of scheme $\Xc^{'}_{1}$, so that for example, instead of sending \eqref{eq:TxX11t2}, we simply send
\begin{align} \label{eq:TxX11t2new}
\xv_{2,t} \!=\! \wv_{2,t} c_{2,t} +  \hat{\hv}^{\bot}_{2,t} ( b_{2,t} + b^{'}_{2,t}   ) + \hat{\gv}^{\bot}_{2,t}  b^{''}_{2,t} +  \hat{\gv}^{\bot}_{2,t} a_{2,t}.
\end{align}
Using these two sub-schemes $\Xc^{'}_{1}$ and $\Xc^{''}_{1}$, one after the other, allows for $\Xc_{1}$ to achieve the symmetric DoF  $d_1=d_2=\frac{ 4+\gamma+ \alpha  }{6}$ for $\lambda_{\gamma\alpha}=\lambda_{\alpha\gamma}=\frac{1}{2}$.

\begin{table}
\caption{CSIT alternating sequence for $\Xc^{''}_{1}$.}
\begin{center}
\begin{tabular}{|c|c|c|c|c|}
  \hline
Phase                               & 1        & 2          & 3          & 4          \\
  \hline
Duration                               & $T_1$        & $T_2$           & $T_3$           & $T_4$           \\
  \hline
$I_1$       & $\gamma$ & $\alpha$ & $\gamma$   & $\alpha$          \\
   \hline
$I_2$        & $\alpha$ & $\gamma$ & $\alpha$   & $\gamma$      \\
   \hline
\end{tabular}
\end{center}
\label{tab:x1half2}
\end{table}
\vspace{3pt}

\begin{example}
For $\gamma=\frac{2}{3}, \alpha=\frac{1}{3}$, the sub-schemes have $T_1=T_2=T_3=T_4=1$ (cf. \eqref{eq:T1T2}), and they jointly achieve DoF point $(\frac{4+\gamma+\alpha}{6}, \frac{4+\gamma+\alpha}{6}) = (\frac{5}{6}, \frac{5}{6})$.
\end{example}

\subsection{Scheme $\Xc_{2}$: achieving DoF points $(1, \frac{\gamma+\alpha}{2})$ and $(\frac{\gamma+\alpha}{2}, 1)$ with $\lambda_{\gamma\alpha}=\lambda_{\alpha\gamma}=\frac{1}{2}$; no delayed CSIT}

This scheme, described for $\lambda_{\gamma\alpha}=\lambda_{\alpha\gamma}=\frac{1}{2}$, will achieve the DoF corner points $(1, \frac{\gamma+\alpha}{2})$ and $(\frac{\gamma+\alpha}{2}, 1)$, and do so without delayed CSIT.
The scheme consists of two channel uses, and it will be described, without loss of generality, for the CSIT alternating sequence $I_1I_2 = \gamma\alpha$ for $t=1$ and $I_1I_2 = \alpha\gamma$ for $t=2$.

During the first channel use $(t=1, I_1=\gamma, I_2=\alpha)$, the transmitter sends
\begin{align} \label{eq:TxX1t1}
\xv_1 = \wv_1 c_1+ \hat{\gv}^{\bot}_1 a_1 + \hat{\gv}^{\bot}_1 a^{'}_1+ \hat{\hv}^{\bot}_1 b_1  \end{align}
($c_1$ common, $a_1,a^{'}_1$ for user 1, $b_1$ for user 2), with
\begin{equation}\label{eq:rpX1t1}
\begin{array}{cccc}
P^{(c)}_1 \doteq P, & P^{(a)}_1 \doteq P^{\gamma}, & P^{(a')}_1 \doteq  P^{\alpha}, & P^{(b)}_1 \doteq P^{\gamma} \\
r^{(c)}_1 = 1-\gamma,   &r^{(a)}_1 = \gamma-\alpha,   & r^{(a')}_1 = \alpha,  & r^{(b)}_1 =\gamma \end{array} \end{equation}
and thus
\begin{eqnarray}
  y^{(1)}_1 \!\!\!\!\!\!\!&=& \!\!\!\!\!\underbrace{\hv^\T_1 \wv_1 c_1}_{P} +\underbrace{\hv^\T_1 \hat{\gv}^{\bot}_1 a_1}_{P^{\gamma}}+ \underbrace{\hv^\T_1 \hat{\gv}^{\bot}_1 a^{'}_1}_{P^{\alpha}} + \underbrace{\tilde{\hv}^\T_1 \hat{\hv}^{\bot}_1 b_1}_{P^{0}}+\underbrace{z^{(1)}_1}_{P^0} \label{eq:x1t1y1}\\
  y^{(2)}_1 \!\!\!\!\!\!\!&=&\!\!\!\!\!\underbrace{\gv^\T_1 \wv_1 c_1}_{P} + \underbrace{\tilde{\gv}^\T_1\hat{\gv}^{\bot}_1 a_1}_{P^{\gamma-\alpha}}+\underbrace{\tilde{\gv}^\T_1\hat{\gv}^{\bot}_1 a^{'}_1}_{P^{0}}  +  \underbrace{\gv^\T_1 \hat{\hv}^{\bot}_1 b_1}_{P^{\gamma}}+\underbrace{z^{(2)}_1}_{P^0}. \label{eq:x1t1y2}
\end{eqnarray}

During the second channel use $(t=2, I_1=\alpha, I_2=\gamma)$, the transmitter sends
\begin{align} \label{eq:TxX1t2}
\xv_2 = \wv_2 c_2+ \hat{\gv}^{\bot}_2 a_2 + \hat{\hv}_2 a_1+ \hat{\hv}^{\bot}_2 b_2  \end{align}
where $a_1$ is the same symbol sent before, and where
\begin{equation}\label{eq:rpX1t2}
\begin{array}{ccc}
P^{(c)}_2 \doteq P,     & P^{(a)}_2 \doteq P^{\gamma}, & P^{(b)}_2 \doteq P^{\alpha} \\
r^{(c)}_2 = 1-\gamma,   & r^{(a)}_2   = \gamma,        & r^{(b)}_2 = \alpha \end{array} \end{equation}
resulting in
\begin{align}
  y^{(1)}_2 \!&=\! \underbrace{\hv^\T_2 \wv_2 c_2}_{P} +\underbrace{\hv^\T_2 \hat{\gv}^{\bot}_2 a_2}_{P^{\gamma}}+ \underbrace{\hv^\T_2 \hat{\hv}_2 a_1}_{P^{\gamma}} + \underbrace{\tilde{\hv}^\T_2 \hat{\hv}^{\bot}_2 b_2}_{P^{0}}+\underbrace{z^{(1)}_2}_{P^0} \label{eq:x1t2y1}\\
  y^{(2)}_2 \!&=\!\underbrace{\gv^\T_2 \wv_2 c_2}_{P} + \underbrace{\tilde{\gv}^\T_2\hat{\gv}^{\bot}_2 a_2}_{P^{0}}+\underbrace{\gv^\T_2\hat{\hv}_2 a_1}_{P^{\gamma}} + \underbrace{\gv^\T_2 \hat{\hv}^{\bot}_2 b_2}_{P^{\alpha}}+\underbrace{z^{(2)}_1}_{P^0}. \label{eq:x1t2y2}
\end{align}

Now we see that both users can decode $c_t,\ t=1,2$ by treating the other signals as noise.
Then user~1 removes $c_1$ from $y^{(1)}_1$ in \eqref{eq:x1t1y1}, successively decodes $a_1$ and $a^{'}_1$, removes $c_2$ and $a_1$ from $y^{(1)}_2$ in \eqref{eq:x1t2y1} and decodes $a_2$.

Similarly user 2, decodes and removes $c_2$ from $y^{(2)}_2$ in \eqref{eq:x1t2y2}, then successively decodes $a_1$ and $b_2$ (recall that $r^{(a)}_1=\gamma-\alpha$ and $r^{(b)}_2 = \alpha$), then removes $c_1$ and $a_1$ from $y^{(2)}_1$ in \eqref{eq:x1t1y2}, and then decodes $b_1$.
Assuming that all information in $c_1, c_2$ is allocated to user~1, gives
\begin{align}
d_1 &= \frac{2(1- \gamma)+ \gamma-\alpha+ \alpha + \gamma}{2} =1 \\
d_2 &= \frac{\gamma+ \alpha }{2}.
\end{align}
Similarly, switching the role of users, and the role of $\alpha$ and $\gamma$, gives the other DoF point $(\frac{\gamma+\alpha}{2}, 1)$.

\subsection{Scheme $\Xc_{3}$: achieving DoF points $(1, \gamma)$ and $(\gamma, 1)$ with $\lambda_{\gamma\gamma}=1$, and DoF points $(1, \alpha)$ and $(\alpha, 1)$ with $\lambda_{\alpha\alpha}=1$; No delayed CSIT}

Let us first consider the case where $\lambda_{\gamma\gamma}=1$.
The scheme consists of one channel use, during which the transmitter sends
\begin{align} \label{eq:TxX3t1}
\xv_1 = \wv_1 c_1+ \hat{\gv}^{\bot}_1 a_1 + \hat{\hv}^{\bot}_1 b_1  \end{align}
($c_1$ common, $a_1$ for user 1, $b_1$ for user 2), with
\begin{equation}\label{eq:rpX3t1}
\begin{array}{ccc}
P^{(c)}_1 \doteq P, & P^{(a)}_1 \doteq P^{\gamma},  & P^{(b)}_1 \doteq P^{\gamma} \\
r^{(c)}_1 = 1-\gamma,   &r^{(a)}_1 = \gamma,  & r^{(b)}_1 =\gamma \end{array} \end{equation}
and as a result
\begin{align}
  y^{(1)}_1 &= \underbrace{\hv^\T_1 \wv_1 c_1}_{P} +\underbrace{\hv^\T_1 \hat{\gv}^{\bot}_1 a_1}_{P^{\gamma}} + \underbrace{\tilde{\hv}^\T_1 \hat{\hv}^{\bot}_1 b_1}_{P^{0}}+\underbrace{z^{(1)}_1}_{P^0} \label{eq:x3t1y1}\\
  y^{(2)}_1 \!&=\underbrace{\gv^\T_1 \wv_1 c_1}_{P} + \underbrace{\tilde{\gv}^\T_1\hat{\gv}^{\bot}_1 a_1}_{P^{0}} +  \underbrace{\gv^\T_1 \hat{\hv}^{\bot}_1 b_1}_{P^{\gamma}}+\underbrace{z^{(2)}_1}_{P^0}. \label{eq:x3t1y2}
\end{align}
User~1 then successively decodes $c_1$ and $a_1$, and user~2 successively decodes $c_1$ and $b_1$.  By assigning $c_1$ entirely to user~1, gives $(d_1=1, d_2=\gamma)$, while assigning $c_1$ to user~2, gives $(d_1=\gamma, d_1=1)$.

Considering the case where $\lambda_{\alpha\alpha}=1$, we simply replace $\gamma$ with $\alpha$, to get DoF points $(1, \alpha)$ and $(\alpha, 1)$, again without delayed CSIT.

\subsection{Merging component schemes and calculating DoF}
We proceed to show the achievability of the DoF regions in Theorem~\ref{thm:alterFre} and Proposition~\ref{prop:alterFreNodelay}, by first showing how the previously described schemes achieve the corner points for any $\lambda_{\gamma\gamma},\lambda_{\alpha\alpha},\lambda_{\gamma\alpha}=\lambda_{\alpha\gamma}$.

To achieve DoF point $( \frac{2+\bar{\lambda}}{3}, \frac{2+\bar{\lambda}}{3})$ - in the presence of delayed CSIT - we combine schemes $\Xc_1,\Xc_4$ and consider communication over a total of $n$ channel uses.  Scheme $\Xc_1$ uses a total of $n(\lambda_{\gamma\alpha}+\lambda_{\alpha\gamma})$ channel uses (for half of which we have $I_1I_2=\gamma\alpha$, else $I_1I_2=\alpha\gamma$) to convey $n (\lambda_{\gamma\alpha}+\lambda_{\alpha\gamma})\frac{4+\gamma+\alpha}{6}\log P$ bits per user.
Then $\Xc_4$ is used for $n\lambda_{\alpha\alpha}$ channel uses (during which $I_1I_2=\alpha\alpha$) to convey $n\lambda_{\alpha\alpha}\frac{2+\alpha}{3}\log P$ bits per user, and then again $\Xc_4$ uses $n\lambda_{\gamma\gamma}$ channel uses (during which $I_1I_2=\gamma\gamma$) to convey $n\lambda_{\gamma\gamma}\frac{2+\gamma}{3}\log P$ bits per user (see Table~\ref{tab:calDoF1}, Table~\ref{tab:calDoF11}).

Consequently
\begin{align}
d_1 =d_2&=  \lambda_{\gamma\gamma} \frac{2+\gamma}{3}+ \lambda_{\alpha\alpha} \frac{2+\alpha}{3} + (\lambda_{\gamma\alpha} +  \lambda_{\alpha\gamma}) \frac{4+\gamma+\alpha}{6}   \nonumber \\
    &=\frac{2  +    \lambda_{\gamma\gamma} \gamma+ \lambda_{\alpha\alpha} \alpha +  (\lambda_{\gamma\alpha} +  \lambda_{\alpha\gamma})\frac{\gamma+\alpha}{2} }{3}   \nonumber \\
	&=\frac{2  +   \bar{\lambda}  }{3}.
\end{align}

\begin{table}
\caption{Component schemes used to achieve DoF $(\frac{2+\bar{\lambda}}{3}, \frac{2+\bar{\lambda}}{3})$}
\begin{center}
\begin{tabular}{|c|c|c|c|c|}
  \hline
Component Scheme       & Setting-CS            & DoF of CS    \\
  \hline
$\Xc_1$          & $\lambda_{\gamma\alpha}=\lambda_{\alpha\gamma}=\frac{1}{2}$   & $(\frac{4+\gamma+\alpha}{6}, \frac{4+\gamma+\alpha}{6})$        \\
  \hline
$\Xc_4$          & $\lambda_{\gamma\gamma}=1$   & $(\frac{2+\gamma}{3}, \frac{2+\gamma}{3})$       \\
   \hline
$\Xc_4$          & $\lambda_{\alpha\alpha}=1$   & $(\frac{2+\alpha}{3}, \frac{2+\alpha}{3})$       \\
   \hline
\end{tabular}
\end{center}
\label{tab:calDoF1}
\end{table}

\begin{table}
\caption{Merging of component schemes to achieve DoF $(\frac{2+\bar{\lambda}}{3}, \frac{2+\bar{\lambda}}{3})$}
\begin{center}
\begin{tabular}{|c|c|c|c|c|}
  \hline
Comp. Scheme      &  Channel uses                &  Bits per user ($\times \log P$)    \\
  \hline
$\Xc_1$          & $n(\lambda_{\gamma\alpha}+\lambda_{\alpha\gamma})$   & $n (\lambda_{\gamma\alpha}+\lambda_{\alpha\gamma})\frac{4+\gamma+\alpha}{6}, $        \\
  \hline
$\Xc_4$          & $n\lambda_{\gamma\gamma}$   & $n\lambda_{\gamma\gamma}\frac{2+\gamma}{3}$       \\
   \hline
$\Xc_4$          & $n\lambda_{\alpha\alpha}$   & $n\lambda_{\alpha\alpha}\frac{2+\alpha}{3}$       \\
   \hline
	& & \\
Sum           & $n$                        & $n\frac{2  +   \bar{\lambda}  }{3}$       \\
   \hline
\end{tabular}
\end{center}
\label{tab:calDoF11}
\end{table}

Similarly, DoF point $( 1,\bar{\lambda})$ can be achieved, without delayed CSIT, for any $\lambda_{\gamma\gamma},\lambda_{\alpha\alpha},\lambda_{\gamma\alpha}=\lambda_{\alpha\gamma}$, by using component schemes $\Xc_2$ and $\Xc_3$ as described in Table~\ref{tab:calDoF2} and Table~\ref{tab:calDoF22}.  Adding up the bits, gives
\begin{align}
d_1 &= 1 \nonumber\\
d_2 &= \lambda_{\gamma\gamma} \gamma + \lambda_{\alpha\alpha} \alpha + (\lambda_{\gamma\alpha} +  \lambda_{\alpha\gamma}) \frac{\gamma+\alpha}{2} =\bar{\lambda}.
\end{align}
Similarly DoF point $(\bar{\lambda}, 1 )$ can also be achieved, again without delayed CSIT.

\begin{table}[h]
\caption{Component schemes used to achieve DoF $( 1,\bar{\lambda})$}
\begin{center}
\begin{tabular}{|c|c|c|c|c|}
  \hline
Component Scheme       & Setting-CS           & DoF of CS    \\
  \hline
$\Xc_2$          & $\lambda_{\gamma\alpha}=\lambda_{\alpha\gamma}=\frac{1}{2}$   & $(1, \frac{\gamma+\alpha}{2})$        \\
  \hline
$\Xc_3$          & $\lambda_{\gamma\gamma}=1$   & $(1, \gamma)$       \\
   \hline
$\Xc_3$          & $\lambda_{\alpha\alpha}=1$   & $(1, \alpha)$       \\
   \hline
\end{tabular}
\end{center}
\label{tab:calDoF2}
\end{table}

\begin{table}[h]
\caption{Achieving DoF $( 1,\bar{\lambda})$, without delayed CSIT}
\begin{center}
\begin{tabular}{|c|c|c|c|c|}
  \hline
Comp. Scheme      &  Channel uses                &  Bits ($\times \log P$)    \\
  \hline
$\Xc_2$          & $n(\lambda_{\gamma\alpha}+\lambda_{\alpha\gamma})$   & $n(\lambda_{\gamma\alpha}+\lambda_{\alpha\gamma}, (\lambda_{\gamma\alpha}+\lambda_{\alpha\gamma})\frac{\gamma+\alpha}{2})$        \\
  \hline
$\Xc_3$          & $n\lambda_{\gamma\gamma}$   & $n(\lambda_{\gamma\gamma}, \lambda_{\gamma\gamma}\gamma)$       \\
   \hline
$\Xc_3$          & $n\lambda_{\alpha\alpha}$   & $n(\lambda_{\alpha\alpha}, \lambda_{\alpha\alpha}\alpha)$       \\
   \hline
	& & \\
Sum          & $n$                        & $n(1, \bar{\lambda}) $       \\
   \hline
\end{tabular}
\end{center}
\label{tab:calDoF22}
\end{table}

Finally the entirety of the optimal DoF region in Theorem~\ref{thm:alterFre} can be achieved by time sharing between the component schemes that achieve DoF corner points $\{ (0,1), (\bar{\lambda},1), (\frac{2+\bar{\lambda}}{3},\frac{2+\bar{\lambda}}{3}),(1,\bar{\lambda}),(1, 0)\}$.  Similarly, the entire DoF region in Proposition~\ref{prop:alterFreNodelay} is achievable, without delayed CSIT, by time sharing between the DoF corner points $\{ (0,1), (\bar{\lambda},1),(1,\bar{\lambda}),(1, 0)\}$.

\section{Conclusions} \label{sec:conclu}
The work has provided the optimal DoF region for the any-two-state alternating CSIT setting in the presence of delayed CSIT.
The corresponding analysis and optimal communication schemes come at a time where it becomes increasingly necessary to communicate in the presence of imperfect timeliness and quality of feedback.


\end{document}